\documentclass[pra,onecolumn,floatfix,superscriptaddress,longbibliography,notitlepage, nofootinbib]{revtex4-1}
\pdfoutput=1

\usepackage[utf8]{inputenc}
\usepackage{braket}
\usepackage{amsmath}

\usepackage{dcolumn}   
\usepackage{bm}        
\usepackage{amssymb}
\usepackage{mathtools}
\usepackage{tikz}
\usepackage{qcircuit}
\usepackage{appendix}
\usepackage{hyperref}
\usepackage{subcaption}
\usepackage{amsmath,amsfonts,amssymb}
\usepackage{mathtools}
\usepackage{thmtools,thm-restate}
\usepackage{algorithm}
\usepackage{mathrsfs}
\usepackage{tikz}
\usepackage{subcaption}
\usepackage{pgfplots}
\usetikzlibrary{3d}
\usepackage{tkz-euclide}

\usepackage{algpseudocode}

\newtheorem{definition}{Definition}

\newtheorem{lemma}{Lemma}

\usetikzlibrary{arrows.meta}

\def\BibTeX{{\rm B\kern-.05em{\sc i\kern-.025em b}\kern-.08em
    T\kern-.1667em\lower.7ex\hbox{E}\kern-.125emX}}

\begin{document}
\title{Improved Time-independent Hamiltonian Simulation }
\author{Nhat A. Nghiem}
\email{nhatanh.nghiemvu@stonybrook.edu}
\affiliation{Department of Physics and Astronomy, State University of New York at Stony Brook, Stony Brook, NY 11794-3800, USA}
\affiliation{C. N. Yang Institute for Theoretical Physics, State University of New York at Stony Brook, Stony Brook, NY 11794-3840, USA}

\begin{abstract}
We describe a simple method for simulating time-independent Hamiltonian $H$ that could be decomposed as $H = \sum_{i=1}^m H_i$ where each $H_i$ can be efficiently simulated. Approaches relying on product formula generally work by splitting the evolution time into segments, and approximate the evolution in each segment by the evolution of composing Hamiltonian $H_i$. This key step incur a constraint, that prohibits a (poly)logarithmic scaling on approximation error. We employ the recently introduced quantum singular value transformation framework to utilize the ability to simulate $H_i$ in an alternative way, which then allows us to construct and simulate the main Hamiltonian $H$ with polylogarithmical scaling on the inverse of desired error, which is a major improvement with respect to product formula approaches.
\end{abstract}
\maketitle

\section{Introduction}
\label{sec: introduction}
Quantum simulation has stand as one of central application of quantum computer. As envisioned by Feynman \cite{feynman2018simulating}, some quantum mechanism is necessary to simulate nature, as nature is essentially quantum. Since then, major progress have been made. Some pioneering works include \cite{lloyd1996universal}, where the author uses Lie-Trotter formula to show that a universal quantum simulator is possible, which means that the ability to simulate some systems can translate directly to the simulation of other systems. In \cite{aharonov2003adiabatic}, the authors show how a system characterized by a sparse Hamiltonian could be efficiently simulated. In a series of works \cite{berry2007efficient, berry2014high, childs2021theory}, methods based on sparse-access Hamiltonian and product formulas have been constructed. Alternatively, the work in  \cite{childs2010relationship, berry2015hamiltonian} uses quantum walk technique developed in \cite{ambainis2003quantum, ambainis2007quantum} to simulate certain Hamiltonian. Model based on linear combination of unitaries have been constructed in \cite{childs2012hamiltonian, berry2015simulating}. Most recently, quantum signal processing was proposed \cite{low2017optimal, low2019hamiltonian} as a very fruitful framework for simulating time-independent Hamiltonian. These aforementioned works involve time-independent regime, and in parallel attempts, simulation of time-dependent Hamiltonian have been constructed \cite{poulin2011quantum, berry2020time, chen2021quantum, an2022time, low2018hamiltonian, kieferova2019simulating}. 

An important aspect of simulation is the complexity, which measures the number of elementary operations, such as single and two qubits gates, required to achieve desired simulation. For approximation purpose, another important metric is the scaling of resources on the desired error. A common technique to simulate given Hamiltonian used in previous simulation results is Lie-Trotter product formula and its higher order extension due to Suzuki \cite{suzuki1991general}. The most thorough analysis of scaling of product formula's approach for simulating Hamiltonian is provided in \cite{childs2021theory}. We remark that while product formula's approach allows a straightforward way to simulate a system, given the ability to simulate ``its constituents'', the key step in such method is breaking the time interval into multiple segments, and in fact, it induces a fundamental constraints to the complexity, as the number of operation need to be executed needs to be as large as the number of such segments. As we shall show subsequently, it leads to the scaling on error term being at least sublinear. 

In this work, motivated by the hurdle of such constraint, we observe an alternative route to remove it. Our method is driven by the recently introduced quantum singular value transformation (QSVT) framework, that allows algebraic quantum operations could be implemented in a simple, effective manner. The essential idea from QSVT is that, if we can build a so-called block encoding of some operator, then we can implement a wide range of function of such operator. As application to quantum simulation, if somehow we can block encode the Hamiltonian $H$ of interest, then we can implement its exponent $\exp(-iHt)$ for some $t$ by making use of the so-called Jacobi-Anger expansion. Building upon such an idea, we would show in this work that the ability to simulate constituting Hamiltonian can be used in a different way, that allows us to construct the block encoding of main Hamiltonian $H$ directly. Hence, it results in a framework with scaling that is (poly)logarithmical in inverse of desired error, which is superpolynomial improvement to the popular product formula approach.

The structure of this paper is as follows. In section \ref{sec: preli}, we summarize some preliminaries, including crucial definitions and technical tools of our work. Section \ref{sec: mainframework} is devoted to our main framework. We beginning in section \ref{sec: mainframework} by describing the product formula, where we would see the fundamental origin of the constraint we mentioned above. Then, in the next part of the same section \ref{sec: mainframework}, we outline our method, which is built from the so-called logarithmic of unitary tool developed in \cite{gilyen2019quantum}. We close and discuss our work in section \ref{sec: conclusion}.

\section{Preliminaries}
\label{sec: preli}
In this section we recapitulate key recipes that would be central to our construction. We refer the readers to original works \cite{gilyen2019quantum} for greater details. 
\begin{definition}
\label{def: blockencode}
    Suppose that $A$ is an s-qubit operator, $\alpha, \epsilon \in \mathbb{R}_+$ and $a \in \mathbb{N}$, then we say that the $(s+a)$-qubit unitary $U$ is an $(\alpha, a ,\epsilon)$-block encoding of $A$, if
    $$ || A - \alpha (\bra{0}^{\otimes a} \otimes \mathbb{I}) U ( \ket{0}^{\otimes a} \otimes \mathbb{I} ) || \leq \epsilon$$
    Equivalently, in matrix representation, $U$ is said to be a block encoding of $A/\alpha$ if $U$ has the form
    \begin{align*}
    U = \begin{pmatrix}
       \frac{A}{\alpha} & \cdot \\
       \cdot & \cdot \\
    \end{pmatrix}.
\end{align*}
where $(.)$ in the above matrix refers to irrelevant blocks that could be non-zero.
\end{definition}

\begin{lemma}[Theorem 2 in \cite{rattew2023non}]
\label{lemma: diagonal}
     Given an n-qubit quantum state specified by a state-preparation-unitary $U$, such that $\ket{\psi}_n=U\ket{0}_n=\sum^{N-1}_{k=0}\psi_k \ket{k}_n$ (with $\psi_k \in \mathbb{C}$ and $N=2^n$), we can prepare an exact block-encoding $U_A$ of the diagonal matrix $A = {\rm diag}(\psi_0, ...,\psi_{N-1})$ with $\mathcal{O}(n)$ circuit depth and a total of $\mathcal{O}(1)$ queries to a controlled-$U$ gate  with $n+3$ ancillary qubits.
\end{lemma}

\begin{lemma}[\cite{camps2020approximate}]
\label{lemma: tensorproduct}
    Given the unitary block encoding $\{U_i\}_{i=1}^m$ of multiple operators $\{M_i\}_{i=1}^m$ (assumed to be exact encoding), then, there is a procedure that produces the unitary block encoding operator of $\bigotimes_{i=1}^m M_i$, which requires a single use of each $\{U_i\}_{i=1}^m$ and $\mathcal{O}(1)$ SWAP gates. 
\end{lemma}

\begin{lemma}[Linear combination of block-encoded matrices]
    Given unitary block encoding of multiple operators $\{M_i\}_{i=1}^m$ and a unitary that prepares the state $\ket{y} = \sum_{i=1}^{m} \sqrt{ sign(y_i) y_i/ \beta } \ket{i}$ where $\beta = \sum_{i=1}^{m} |y_i|$. Then, there is a procedure that produces a unitary block encoding operator of \,$\sum_{i=1}^m  y_i M_i/ \beta $ in complexity $\mathcal{O}(m)$, using block encoding of each operator $M_i$ a single time. 
    \label{lemma: sumencoding}
\end{lemma}

\begin{lemma}[Block Encoding of Product of Two Matrices]
\label{lemma: product}
    Given the unitary block encoding of two matrices $A_1$ and $A_2$ (assuming to have norm less than 1), then there exists an efficient procedure that constructs a unitary block encoding of $A_1 A_2$ using each block encoding of $A_1,A_2$ one time. 
\end{lemma}

\begin{lemma}\label{lemma: amp_amp}[\cite{gilyen2019quantum} Theorem 30]
\label{lemma: amplification}
Let $U$, $\Pi$, $\widetilde{\Pi} \in {\rm End}(\mathcal{H}_U)$ be linear operators on $\mathcal{H}_U$ such that $U$ is a unitary, and $\Pi$, $\widetilde{\Pi}$ are orthogonal projectors. 
Let $\gamma>1$ and $\delta,\epsilon \in (0,\frac{1}{2})$. 
Suppose that $\widetilde{\Pi}U\Pi=W \Sigma V^\dagger=\sum_{i}\varsigma_i\ket{w_i}\bra{v_i}$ is a singular value decomposition. 
Then there is an $m= \mathcal{O} \Big(\frac{\gamma}{\delta}
\log \left(\frac{\gamma}{\epsilon} \right)\Big)$ and an efficiently computable $\Phi\in\mathbb{R}^m$ such that
\begin{equation}
\left(\bra{+}\otimes\widetilde{\Pi}_{\leq\frac{1-\delta}{\gamma}}\right)U_\Phi \left(\ket{+}\otimes\Pi_{\leq\frac{1-\delta}{\gamma}}\right)=\sum_{i\colon\varsigma_i\leq \frac{1-\delta}{\gamma} }\tilde{\varsigma}_i\ket{w_i}\bra{v_i} , \text{ where } \Big|\!\Big|\frac{\tilde{\varsigma}_i}{\gamma\varsigma_i}-1 \Big|\!\Big|\leq \epsilon.
\end{equation}
Moreover, $U_\Phi$ can be implemented using a single ancilla qubit with $m$ uses of $U$ and $U^\dagger$, $m$ uses of C$_\Pi$NOT and $m$ uses of C$_{\widetilde{\Pi}}$NOT gates and $m$ single qubit gates.
Here,
\begin{itemize}
\item C$_\Pi$NOT$:=X \otimes \Pi + I \otimes (I - \Pi)$ and a similar definition for C$_{\widetilde{\Pi}}$NOT; see Definition 2 in \cite{gilyen2019quantum},
\item $U_\Phi$: alternating phase modulation sequence; see Definition 15 in \cite{gilyen2019quantum},
\item $\Pi_{\leq \delta}$, $\widetilde{\Pi}_{\leq \delta}$: singular value threshold projectors; see Definition 24 in \cite{gilyen2019quantum}.
\end{itemize}
\end{lemma}

\section{Main Framework}
\label{sec: mainframework}
\begin{center}
    \textit{Review of Product Formula }
\end{center}
Let the Hamiltonian of interest $H$ be $H = \sum_{i=1}^m H_i$ where each $H_i$ is a time-independent and that it can be efficiently simulated. For example, it can be done by using sparse access method \cite{berry2007efficient}, quantum walk \cite{childs2010relationship, berry2015hamiltonian}, quantum signal processing \cite{low2017optimal}, etc. The basic idea of product formula is to split the time evolution of $H$ into simpler evolutions for (sufficiently) small time steps. The most general theory has been developed in \cite{childs2021theory}. Using recursively high-order Suzuki's technique \cite{suzuki1991general}, we have the following \cite{childs2021high, berry2007efficient}: 
\begin{lemma}
    Let $H=\sum_{i=1}^m H_i$. Define $S_2(\lambda) = \prod_{i=1}^m e^{H_i \lambda/2} \prod_{i'=m}^1 e^{H_i' \lambda/2} $, and recursion relation:
    $$ S_{2k} (\lambda)  = [S_{2k-2}(p_k\lambda]^2   S_{2k-2} ( (1-4p_k)\lambda  ) [S_{2k-2}(p_k\lambda]^2$$
    with $p_k = (4- 4^{1/(2k-1)})^{-1}$ for $k>1$. Then we have that:
    \begin{align}
        || \exp(-i Ht) - S_{2k}( -it) || \leq 2 \Gamma^{p+1} \frac{t^{p+1}}{p+1} \alpha^{(p)}
    \end{align}
    where $\Gamma = 2 \times 5^{k-1}$ and 
    $$\alpha^{(p)} = \sum_{i,j,..., 2k+1= 1 }^m ||  [H_{2k+1} [H_{2k}, ...[H_2,H_1]]  ]  || $$
    Furthermore, dividing the time interval $[0,t]$ into $r$ segments, we have: 
    \begin{align}
        || \exp(-iHt) - S_{2k}(-it/r)^r || \leq r || \exp(-i H\frac{t}{r} - S_{2k} (\frac{-it}{r})  || \leq  2 \Gamma^{2k+1} \frac{t^{p+1}}{r^{2k} (2k+1)} \alpha^{(2k)}
    \end{align}
\end{lemma}
The above result suggests that, in order to have a total error to be $\delta$, we need:
\begin{align}
    2 \Gamma^{2k+1} \frac{t^{p+1}}{r^{2k} (2k+1)} \alpha^{(2k)} = \delta
\end{align}
which implies that:
\begin{align}
    r = \mathcal{O}( \frac{1}{\delta^{1/2k}} )
\end{align}
Thus, the complexity of simulating $H$ is (at best) sublinear in $1/\delta$, as we need to use the simulation of each composing Hamiltonian $H_i$ at least $\mathcal{O}(r)$ times. Hence, the product formula itself possess a fundamental constraint on the (inverse of) error scaling, which seems inprohibitive. Now we proceed to describe a method for improving such scaling to polylogarithmic in $1/\delta$.  \\

\begin{center}
    \textit{Improved Framework}
\end{center}
Let $\exp(-i H_i t)$ be the evolution operator being simulated at time $t$. As it is a unitary operator, it block encodes itself (see further \cite{gilyen2019quantum} and appendix \ref{def: blockencode} for reference). Our framework is based on the following result, which is Corollary 71 from \cite{gilyen2019quantum}:
\begin{lemma}
\label{lemma: logarithmicofunitary}
    Suppose that $U = exp(-iH)$, where $H$ is a Hamiltonian of (spectral) norm $||H|| \leq 1/2$. Let $\epsilon \in (0,1/2]$ then we can implement a $(2/\pi, 2,\epsilon)$-block encoding of $H$ with $\mathcal{O}( \log(1/\epsilon)$ uses of controlled-U and its inverse, using $\mathcal{O}(\log(1/\epsilon)$ two-qubits gates and using a single ancilla qubit. 
\end{lemma}
The above result, combined with the ability to simulate $\exp(-i H_i)$ (basically simulating $H_i$ up to $t=1$), allows us to obtain the block encoding of $\pi H_i/2$. Let $T_i$ be the complexity (could be query, or the gate complexity, i.e, the number of queries or gates required) of obtaining $\exp(-iH_i)$. Then the total complexity of obtaining the block encoding of $\pi H_i/2$ is $\mathcal{O}(  T_i \log \frac{1}{\epsilon} )$ (with extra 2 more qubits). 

Given the block encodings of $\pi H_i/2$ (for $i=1,2,...,m$), the technique developed in Lemma 52 of \cite{gilyen2019quantum} (see lemma \ref{lemma: sumencoding} of appendix) allows us to use the block encoding of each $\pi H_i/2m$ to construct the block encoding of:
\begin{align}
    \frac{1}{m } \sum_{i=1}^m \frac{\pi}{2} H_i = \frac{\pi H}{2m}
\end{align}
Then again, one can use quantum singular value transformation technique \cite{gilyen2019quantum} to simulate $ \exp(- i\frac{\pi H}{2m} t) $. More thoroughly, the central result of quantum singular value transformation framework is:
\begin{lemma}\label{lemma: qsvt}[\cite{gilyen2019quantum} Theorem 56]
\label{lemma: theorem56}
Suppose that $U$ is an
$(\alpha, a, \epsilon)$-encoding of a Hermitian matrix $A$. (See Definition 43 of \cite{gilyen2019quantum} for the definition.)
If $P \in \mathbb{R}[x]$ is a degree-$d$ polynomial satisfying that
\begin{itemize}
\item for all $x \in [-1,1]$: $|P(x)| \leq \frac{1}{2}$.
\end{itemize}
Then, there is a quantum circuit $\tilde{U}$, which is an $(1,a+2,4d \sqrt{\frac{\epsilon}{\alpha}})$-encoding of $P(A/\alpha)$, and
consists of $d$ applications of $U$ and $U^\dagger$ gates, a single application of controlled-$U$ and $\mathcal{O}((a+1)d)$
other one- and two-qubit gates.
\end{lemma}
To apply the above lemma, we make use of Jacobi-Anger expansion \cite{low2017optimal, low2019hamiltonian, childs2017lecture} to approximate $\exp(-i x t)$ (for some real $x$) by some polynomials, i.e.,
\begin{align}
    \exp(-i xt) \approx J_0(-t) + 2 \sum_{k=1}^K i^k J_k(-t) T_k(x)
\end{align}
where $J_k$ is Bessel function and $T_k$ is Chebyshev polynomial. To obtain good approximation, for example, to make the approximation $\mathcal{O}(\epsilon)$, the value of $K$ to be $\mathcal{O}(t + \log(\frac{1}{\epsilon} )/ \log( e + \ln(\frac{1}{\epsilon})/t) $. Hence, these polynomials could be implemented using QSVT result \cite{gilyen2019quantum} in a simple manner, as summarized in the above lemma. 

Thus, it is straightforward to obtain the block encoding of $\exp(-i \frac{\pi H}{2m}t )$ for some fixed $t$. One can then note that, by scaling $t \longrightarrow 2t/\pi$, and repeating $m$ applications:
\begin{align}
    \prod_{1}^m \exp(-i \frac{\pi H}{2m} \frac{2t}{\pi} ) = \exp(-i Ht)
\end{align}

Now we analyze the error resulting from approximating the block encoding of individual $H_i$. Remind that using lemma \ref{lemma: logarithmicofunitary} resulted in the block encoding of $\epsilon$-approximated operator to $\pi H_i/2$. Hence, the error accumulated when constructing combination $\frac{1}{m} \sum_{i=1}^m \frac{\pi H_i}{2}$ is $m\epsilon$. Hence, according to lemma \ref{lemma: theorem56}, the error resulted in for the approximation of block encoding of $\exp(-i \pi H t/2m)$ is $\mathcal{O}( 4(t + \log(\frac{1}{\epsilon}) ) \sqrt{m\epsilon} )$. The error further accumulated in the products: $ \prod_{1}^m \exp(-i \frac{\pi H}{2m} \frac{2t}{\pi} )$, led to the error of approximating $\exp(-iHt)$ is $\mathcal{O}( 4(t + \log(\frac{1}{\epsilon}) ) \sqrt{m\epsilon} m )$. To make this error $\delta$, we need to set:
\begin{align}
    4( t+ \log \frac{1}{\epsilon}) \sqrt{ m \epsilon} m = \delta\\
    \longrightarrow \epsilon = \frac{\delta^2}{ 16 (t+ \log \frac{1}{\epsilon} )^2 m^3 }
\end{align}

Given such $\epsilon$, the total complexity for simulating $\exp(-i Ht)$ up to error $\delta$ is then: 
$$ \mathcal{O}( m^2 (\sum_{i=1}^m T_i) \log ( \frac{16(t + \log (1/\epsilon))^2 m^3}{\delta^2    } )  (t + \log( \log (\frac{1}{\delta^2}) ) )  )$$
which is logarithmical dependence on inverse of error. We remark that $T_i$ is the complexity of simulating $H_i$ for time $t=1$, for which optimal result have been founded \cite{low2017optimal}, which is
$$\mathcal{O}( d_i ||H_i|| + \log \frac{1}{\epsilon} )$$
where $d_i$ is the sparsity of $H_i$ and $||H_i||$ is the spectral norm. As we have assumed the ability to simulate such constituent Hamiltonian, defining $d_{\max} = \max_i \{d_i\}$ and $||H||_{\max} = \max_i \{ ||H_i || \}$, then we have that $\sum_{i} T_i = \mathcal{O}( m (d_{\max} ||H||_{\max} + \log(1/\delta) )$. Hence, the total complexity is:
$$ \mathcal{O}( m^3 (d_{\max} ||H||_{\max} + \log \frac{1}{\delta} )\log ( \frac{16(t + \log (1/\delta))^2 m^3}{\delta^2    } )  (t + \log( \log (\frac{1}{\delta^2}) ) )  )$$
Comparing to the standard product formula approach, our framework achieves polylogarithmcal dependence on inverse of error tolerance, which is superpolynomial improvement. 

\section{Conclusion}
\label{sec: conclusion}
We have provided a simple method for constructing evolution operator of some Hamiltonian that could be written as summation of efficiently simulable Hamiltonian. Our method is motivated by the key constraint possessed by the standard product formula, e.g., Lie-Trotter-Suzuki formula, which prohibits logarithmical scaling on inverse of error tolerance. Thanks to the recently introduced quantum singular value transformation framework, we are able to use the simulation ability of composing Hamiltonian in an indirect manner, to construct the desired Hamiltonian directly. Then it can be efficiently simulated again, using technique from quantum singular value transformation framework.

\subsection*{Acknowledgement}
We acknowledge support from Center for Distributed Quantum Processing, Stony Brook University.

\bibliography{ref.bib}{}
\bibliographystyle{unsrt}

\clearpage
\newpage
\onecolumngrid
\appendix

\end{document}